\documentstyle[12pt]{article}
\begin{document}

\begin{titlepage}

\title{Two regularizations -- two different models  of Nambu--Jona-Lasinio}

\author{ R.G. Jafarov\footnote{Baku State University, Baku,
Azerbaijan}  and
 V.E. Rochev \footnote{Institute for High Energy Physics, Protvino, Russia
 (e-mail: rochev@mx.ihep.su)} }
\date{}

\end{titlepage}
\maketitle
\begin{abstract}
Two variants of the Nambu--Jona-Lasinio  model -- the model with
4-dimensional cutoff and the  model with dimensionally-analytical
regularization -- are systematically compared. It is shown that
they are, in essence, two different models of  light-quark
interaction. In the mean-field approximation the distinction
becomes apparent in a behavior of scalar amplitude near the
threshold. For 4-dimensional cutoff  the pole term can be
extracted, which corresponds to  sigma-meson. For
dimensionally-analytical regularization the singularity of the
scalar amplitude is not pole, and  this singularity is quite
disappeared at some value of the regularization parameter.

Still more essential distinction of these models  exists in the
next-to-leading order of  mean-field expansion. The calculations
of meson contributions in the quark chiral condensate and in the
dynamical quark mass  demonstrate, that these contributions though
their relatively smallness can destabilize the Nambu--Jona-Lasinio
model with 4-dimensional cutoff. On the contrary,  the
Nambu--Jona-Lasinio model with dimensionally-analytical
regularization is stabilized with the next-to-leading order, i.e.
 the value of the regularization parameter shifts to the
stability region, where these contributions decrease.
\end{abstract}

\newcommand{\ba}{\begin{eqnarray}}
\newcommand{\ea}{\end{eqnarray}}
\newcommand{\tr}{\,\mbox{tr}\,}

\newpage

\section*{Introduction}

The Namby--Jona-Lasinio (NJL) model \cite{NJL} with the quark
content \cite{QNJL} is one of the most successful effective models
of quantum chromodynamics  of light hadrons in the
non-perturbative region (see, for example, reviews \cite{kle} and
\cite{HatKun} and references therein).

Since the foundation of the NJL model is a non-renormalizable
interaction, the quite essential point of the model is a
regularization.  It already advances in the literature an opinion,
that the NJL model for different regularization can lead to
different physical results. But as concerning to most common
regularizations (such, for example, the 4-dimensional cutoff in
comparison with the Fock--Schwinger "proper-time" regularization
or the Pauli--Villars regularization) this statement is not mean
some principal distinctions of main effects in the leading
approximation of the model. In the next-to-leading order, which
includes the meson contributions in  chiral condensate and
corrections to the quark propagator, these distinctions become
apparent more clearly (see, for example, \cite{Nikolov} -
\cite{Oertel}), but do not change essentially the physical content
of the model in the case too.

Nevertheless, a regularization of the NJL model exists in which
the physical effects differ from the effects of the classical
variant of the model with 4-dimensional cutoff as early as on the
level of two-particle amplitudes. It is a dimensional
regularization considered as a variant of the analytical
regularization. Note, the traditional treatment of the dimensional
regularization as a transition to D-dimensional space strikes in
the application to the NJL model the  essential obstacle: the
regularization parameter, i.e. a deviation in physical dimension
of space, is included in formulae for physical quantities. This
circumstance makes an interpretation of results to be very
awkward. In the alternative treatment of dimensional
regularization as a variant of analytical regularization all
calculations are made in four-dimensional Euclidean space, and the
regularization parameter is treated as a power of a weight
function, which regularizes divergent integrals\footnote{We shall
refer to this regularization as dimensionally-analytical
regularization.  In applications to renormalizable models this
variant of dimensional regularization leads to  same results as
the usual treatment.}. Such treatment of dimensional
regularization was consistently developed for the NJL model in
mean-field approximation by Krewald and Nakayama
 \cite{Krewald}. In work \cite{JaRo} in the framework of this
 regularization the meson contributions in chiral condensate were
 calculated. It should be stressed that in such treatment of
 dimensional regularization the regularization parameter is not a
 deviation in the physical dimension of space. A possible
 treatment of this parameter (see \cite{JaRo}) is a measure of
 some effective gluon influence  on four-fermion quark
 self-action of NJL model.

In this work we make a systematical comparison of the
dimensionally-analytically  regularized NJL model with the
classical variant of NJL model with 4-dimensional
cutoff\footnote{We consider here the SU(2)-NJL model, i.e. the
NJL model with chiral $SU_V(2)\times SU_A(2)$-symmetry.}.

In Sections 1 and 2 the leading-approximation results for the
chiral condensate and two-particle amplitudes are given. At this
point the distinction becomes apparent in scalar amplitude. For
4-dimensional cutoff and for other similar regularizations the
scalar amplitude holds the singularity of pole type, which is
usually interpreted as a sigma-meson with  mass $2m$, where $m$ is
a  quark mass. But for the dimensionally-analytical regularization
the singularity of scalar amplitude is not a pole (moreover, for
some value of the regularization parameter this singularity quite
disappears), and the particle interpretation seems to be
inconsistent. Note, that the interpretation of singularity of
scalar amplitude in the NJL model as a particle meets with
 the well-known difficulties in a comparison with the physical
 spectrum of scalar-meson resonances (see, for example, \cite{Ochs}).

A main distinction becomes apparent in the next-to-leading order
of mean-field expansion, therefore the main results are
calculations made in this order in Section 3. Apart from the
corrections to chiral condensate we  calculate also the
corrections to quark mass for both regularizations. At that, in
contrast to work \cite{JaRo} in which for the scalar amplitude
were used a pole approximation, we use here more exact
leading-singularity approximation. For both regularizations the
leading contribution in first-order chiral condensate is the
contribution of pseudoscalar  (pion) amplitude. But for the
dimensionally-analytical regularization this contribution has the
same sign as the leading-order condensate, and for 4-dimensional
cutoff it has the opposite sign. This distinction is a determining
factor for the problem of stability of the model with respect to
quantum fluctuations caused by the meson amplitudes. In Section 4
a fixation of the model parameters are made with the taking into
account the meson corrections. It is shown, that the coincidence
of sign of the meson contributions with the sign of leading
contribution in the model with dimensionally-analytical
regularization ensures the stability of the model with respect to
quantum fluctuations. Contrary, for the model with 4-dimensional
cutoff these corrections have the opposite sign and can lead to a
destabilization. This destabilization means that an existence  of
the set of  model parameters  critically depends on the value of
chiral condensate $c$: at $\vert c\vert \leq 230$~MeV a system of
equations for the model parameters has no solution. Therefore the
SU(2)-NJL model with 4-dimensional cutoff seems to be in a
dangerous zone of the non-stability with respect to quantum
fluctuations. A simple estimate demonstrates that for SU(3)-model
a situation can merely take a turn for the worse. This result in a
sense has something in common with the statement of work
\cite{Kleinert} in which an applicability of the NJL model to the
description of phenomenon of dynamical  chiral symmetry breaking
(DCSB) was called  in question. (See also a discussion of this
problem in works \cite{Oertel, Ripka}.) Therefore we can maintain
that the NJL model with dimensionally-analytical regularization
and the NJL model with 4-dimensional cutoff are, in essence, two
different models of the light-quark interaction in
non-perturbative region.

\section{Leading order and chiral condensate}

 We consider the NJL model with Lagrangian
\begin{equation}
{\cal L}=\bar \psi i\hat \partial\psi+\frac{g}{2}
\biggl[(\bar\psi\psi)^2+(\bar\psi i\gamma_5 \tau^a\psi)^2\biggr].
\label{LSU2}
\end{equation}
Here $\psi\equiv\psi^{\alpha, c}_j$, where $\alpha=1, 2, 3, 4$ is
Dirac spinor index, $c=1,\ldots, n_c$ is colour index, $j=1,2$ is
isotopic (flavour) index, $\tau^a$ are generators of $SU(2)$-group
(Pauli matrices), $a= 1, 2, 3$. This model possesses the chiral
symmetry of $SU_V(2)\times SU_A(2)$-group.

The mean-field expansion in  bilocal-source formalism \cite{Ro1}
for this model is constructed with the scheme which is given in
work \cite{JaRo}.

In the leading approximation the unique connected Green function
is  quark propagator
\begin{equation}
S^{(0)}_{cd, jk}= \delta_{cd}\delta_{jk}(m-\hat p)^{-1},
\label{S0}
\end{equation}
where dynamical quark mass   $m$ is a solution of  gap equation
\begin{equation}
1= -8ign_c\int\frac{d\tilde q}{m^2-q^2}. \label{gap}
\end{equation}
(Here and below we include a phase factor in an integration over
 momentum space: $d\tilde q\equiv d^4q/(2\pi)^4$.)

The basic order parameter, which defines a degree of DCSB, is a
quantity
$$
\chi=<0\vert\bar\psi\psi\vert 0>=i\tr S(x)\vert_{x\rightarrow0},
$$
where the trace is taken over all discrete indices. It is easy to
see, that from equations (\ref{S0}) and (\ref{gap}) it follows
\begin{equation}
\chi^{(0)}=i\tr S^{(0)}(x)\vert_{x\rightarrow0}=-\frac{m}{g}.
\label{chi0}
\end{equation}
It is a regularization-independent formula.

Quark chiral condensate $c$ is defined for each flavour
separately. In the chiral limit  it is
\begin{equation} c=\biggl(\frac{\chi}{2}\biggr)^{1/3}.\label{cond}
\end{equation}

The integral in equation (\ref{gap}) is divergent, and it should
be considered as a regularization. In Euclidean momentum space
equation(\ref{gap}) has the form
$$
1=\frac{gn_c}{2\pi^2}\int \frac{q^2_e dq^2_e}{m^2+q^2_e}.
$$
Let introduce in the integrand a weight function $w(q^2_e)$, which
will define a choice of regularization. For 4-dimensional cutoff
the weight function is chosen in the form
\begin{equation}
 w_\Lambda(q^2_e)= \theta(\Lambda^2-q^2_e),
\label{cut}
\end{equation}
and equation (\ref{gap}) is
\begin{equation}
1=\kappa_\Lambda\biggr(1-\frac{m^2}{\Lambda^2}
\log(1+\frac{\Lambda^2}{m^2})\biggl), \label{gapcut}
\end{equation}
where
 $\kappa_\Lambda=gn_c\Lambda^2/2\pi^2$.
This equation exactly correspond to the classical result of work
\cite{NJL}.

For dimensionally-analytical regularization the weight function is
chosen as
\begin{equation}
w_\xi(q^2_e)=\frac{1}{\Gamma(1-\xi)}\biggl( \frac{
M^2}{q^2_e}\biggl)^{1+\xi}. \label{AR}
\end{equation}
Integral over  $dq^2_e$ from 0 to $\infty$ converge at $0<\xi<1$,
and equation (\ref{cut}) for dimensionally-analytical
regularization has the form:
\begin{equation}
1=\kappa\Gamma(\xi)\Biggr(\frac{ M^2} {m^2}\Biggl)^{1+\xi},
\label{gapAR}
\end{equation}
where $\kappa=gn_cm^2/2\pi^2$. Factor $\frac{1}{\Gamma(1-\xi)}$
 in weight function (\ref{AR}) are introduced for
 the   correspondence of integration results to that
of the usual prescription of the dimensional regularization as a
formal transition in $D$-dimensional space.  Note, that parameter
$\xi$ differs from the commonly used parameter $\varepsilon$,
which is defined by relation $D=4-2\varepsilon$. As easy to see,
these parameters are connected by relation $\varepsilon=1+\xi$.
Introduction  of this new notation prevents unnecessary
associations with the usual treatment of dimensional
regularization. Furthermore, in terms of the parameter $\xi$ all
subsequent formulae of the NJL model acquire most simple form.

\section{Two-particle amplitude and model parameters
 in leading approximation}

  First-step two-particle amplitude  $A$
(a connected part of  amputated two-particle function) possesses
the following colour and flavour structure:
\begin{equation}
 A^{cd, jk}_{c'd',
j'k'}
=\delta^{cd}\delta^{c'd'}\biggl[\delta_{jk}\delta_{j'k'}A_\sigma
 + \tau^a_{jk}\tau^a_{j'k'}
A_\pi \biggr]. \label{ASU2}
\end{equation}
Here $A_\sigma$ is the scalar amplitude, and   $A_\pi$ is the
pseudoscalar amplitude. In momentum space these amplitudes of the
NJL model depend on a momentum $p$ only, where $p$ is a sum of
quark and antiquark momenta. The amplitudes have the form
\cite{JaRo}:
\begin{equation}
A_\sigma(p)=-\frac{ig}{1-L_S(p)}, \label{Asigma}
\end{equation}
 where $ L_S(p)=ig\int
d\tilde q \; \tr S^{(0)}(p+q)S^{(0)}(q) $ is the scalar quark
loop, and
\begin{equation}
A_\pi(p)=\frac{ig}{1+L_P(p)}, \label{Api}
\end{equation}
where $ L_P(p)=ig\int d\tilde q \;\tr
S^{(0)}(p+q)\gamma_5S^{(0)}(q)\gamma_5 $ is the pseudoscalar quark
loop.

Using identities
$$
\frac{m^2+q^2+(pq)}{(m^2-(p+q)^2)(m^2-q^2)}=\frac{1}{2}\Biggl(
-\frac{1}{m^2-q^2}-\frac{1}{m^2-(p+q)^2}
+\frac{4m^2-p^2}{(m^2-(p+q)^2)(m^2-q^2)}\Biggr),
$$
$$
\frac{m^2-q^2+(pq)}{(m^2-(p-q)^2)(m^2-q^2)}=\frac{1}{2}\Biggl(
\frac{1}{m^2-q^2}+\frac{1}{m^2-(p-q)^2}
+\frac{p^2}{(m^2-(p-q)^2)(m^2-q^2)}\Biggr)
$$
and gap equation (\ref{gap}), it is easy to obtain for $A_\sigma$
and $A_\pi$ the following representations:
\begin{equation}
A_\sigma(p)=\frac{1}{4n_cI_0(p^2)(4m^2-p^2)} \label{A_sigma},
\end{equation}
\begin{equation}
 A_\pi(p)=\frac{1}{4n_cI_0(p^2)p^2}.
\label{A_pi}
\end{equation}
Here
\begin{equation}
 I_0(p^2 )=\int d\tilde
q\frac{1}{(m^2-(p+q)^2)(m^2-q^2)}. \label{I0}
\end{equation}

Integral $I_0$ is calculated as above. Transforming to Euclidean
metric, introducing a standard Feynman parameterization, and
changing an integration variable (which is possible due to
translational invariance of the procedure, see \cite{Krewald}), we
can perform the angular integration. According to the our rules,
then we introduce into the integrand a weight function (\ref{cut})
(for 4-dimensional cutoff), or (\ref{AR}) (for
dimensionally-analytical regularization), and calculate the
integral over  $dq^2_e$. For dimensionally-analytical
regularization ({\it DAR}) we again obtain the result, which
 corresponds to the result of integration with the formal
transition to $D$-dimensional space:
$$
I_0^{DAR}(p^2)= \frac{i\Gamma(1+\xi)}{(4\pi)^2} \int_0^1
du\Biggl(\frac{M^2}{m^2-u(1-u)p^2}\Biggr)^{1+\xi}.
$$
The integral over $dq^2_e$ converges at $-1<\xi<1$. Taking into
account  gap equation (\ref{gapAR}) we obtain:
\begin{equation}
I_0^{DAR}(p^2)=\frac{i}{(4\pi)^2} \frac{\xi}{\kappa}\int_0^1
du\Biggl(1-u(1-u)\frac{p^2}{m^2}\Biggr)^{-1-\xi}=
\frac{i}{(4\pi)^2} \frac{\xi}{\kappa}F(1+\xi, 1; 3/2;
\frac{p^2}{4m^2}), \label{I0AR}
\end{equation}
where $F(a, b; c; z)$ is Gauss hypergeometric function.

For 4-dimensional cutoff ({\it FDC}) we correspondingly obtain:
\begin{equation}
I_0^{FDC}(p^2)=\frac{i}{(4\pi)^2}\int_0^1
du\Bigl[\log\Bigl(1+\frac{\Lambda^2}
{m^2-u(1-u)p^2}\Bigr)-\frac{\Lambda^2}{\Lambda^2+m^2-u(1-u)p^2}
\;\Bigr]. \label{I0cut}
\end{equation}

Formulae for the condensate and the two-particle amplitudes allow
to fix values of the model parameters in the leading approximation
of mean-field expansion. For this purpose we use
regularization-independent formulae (\ref{chi0}), (\ref{cond}) and
a formula for pion decay constant in the NJL model (see
\cite{kle}):
\begin{equation}
f^2_\pi=-4in_cm^2I_0(0) \label{fNJL}.
\end{equation}
For dimensionally-analytical regularization we obtain from
(\ref{I0AR}):
\begin{equation}
I_0^{DAR}(0)= \frac{i}{(4\pi)^2} \frac{\xi}{\kappa},
\label{I0ARzero}
\end{equation}
and for 4-dimensional cutoff (see (\ref{I0cut})):
\begin{equation}
I_0^{FDC}(0)=
\frac{i}{(4\pi)^2}\Bigl[\log\frac{\Lambda^2+m^2}{m^2}-
\frac{\Lambda^2}{\Lambda^2+m^2}\Bigr].
 \label{I0cutzero}
\end{equation}
Correspondingly we obtain for dimensionally-analytical
regularization very simple formula:
\begin{equation}
(f_\pi^2)^{DAR}=\frac{\xi}{2g} \label{fAR}.
\end{equation}
For 4-dimensional cutoff the analogous formula is
\begin{equation}
(f^2_\pi)^{FDC}=\frac{3m^2}{4\pi^2}
\Bigl[\log(1+\frac{\Lambda^2}{m^2})-\frac{\Lambda^2}{m^2+\Lambda^2}\Bigr].
\label{fcut}
\end{equation}

These formulae together with formulae (\ref{chi0})-(\ref{cond})
for condensate and gap equation (equation (\ref{gapAR}) for
dimensionally-analytical regularization and equation
(\ref{gapcut}) for 4-dimensional cutoff) allow to define the values
of principal model parameters.

For pion decay constant we choose the value  $f_\pi=93$ MeV.
Chiral quark condensate $c$ is not directly measured value, and we
shall determine  sets of parameters for some  typical values of
this quantity. For dimensionally-analytical regularization it is
necessary also to fix a value of $M$ ("subtraction point"). In
work \cite{JaRo} we have used for this purpose a value of decay
width $\pi^0\rightarrow 2\gamma$. Analysis of results of this work
demonstrates, that for very large range of condensate values the
value of $M$ is practically permanent and coincides with the value
of dynamical quark mass: $M\approx m$.  Since here we shall take
$M=m$. Such fixation of $M$ equalizes the parameter number   of
dimensionally-analytical regularization  with that of other
regularizations. Gap equation (\ref{gapAR}) with such fixation of
$M$ takes on a very simple form
\begin{equation}
1=\kappa\Gamma(\xi). \label{gapARM}
\end{equation}

The results of parameter fixing in the leading approximation (at
$n_c=3$) are given in Table 1 (dimensionally-analytical
regularization) and in Table 2 (4-dimensional cutoff).

\begin{center}
\begin{tabular}{|l|l|l|l|}\hline
 $c$ (MeV) & $m$ (MeV) & $\xi$ & $\kappa=3gm^2/2\pi^2$\\
 \hline
\  -210  & \ 357  & \  0.333  & \   0.373 \\
\  -220  & \ 356 &  \  0.289 & \   0.322 \\
\  -230  & \ 354 &  \  0.252 & \   0.277 \\
\  -240 &  \ 353 &  \  0.221 & \   0.242 \\
\  -250 &  \ 352  &  \ 0.195  & \  0.212 \\

 \hline

\end{tabular}
\vspace{1cm}

{\small Table 1. The model parameters in  leading order
(dimensionally-analytical regularization): chiral condensate
 $c$, quark mass $m$, regularization parameter $\xi$
 and dimensionless coupling  $\kappa$.}
\end{center}
\vspace{1cm}
\begin{center}
\begin{tabular}{|l|l|l|l|}\hline
 $c$ (MeV)& $m$ (MeV)& $\Lambda$ (MeV)& $\kappa_\Lambda=3g\Lambda^2/2\pi^2$\\
 \hline
\  -210  & \ 423  & \  733  & \   1.86 \\
\  -220  & \ 323 &  \  791 & \   1.448 \\
\  -230  & \ 276 &  \  873 & \   1.315 \\
\  -240 &  \ 253 &  \  947 & \   1.240 \\
\  -250 &  \ 236  &  \ 1029   & \  1.187 \\

 \hline

\end{tabular}
\vspace{1cm}

{\small Table 2. The model parameters in  leading order
(4-dimensional cutoff): chiral condensate
 $c$, quark mass $m$, regularization parameter  $\Lambda$
 and dimensionless coupling  $\kappa_\Lambda$.}
\end{center}

\vspace{1cm}

As it is seen from these Tables the value of the main parameter --
quark mass $m$ -- in the model with 4-dimensional cutoff
is much more sensitive to the value of chiral condensate in
comparison with that of the model with dimensionally-analytical
regularization. At the same time it is necessary to point, that
there  are no some principal distinctions of these variants of the
NJL model at the level of leading approximation for quark
propagator and two-particle amplitudes with the exception of a
behavior of scalar amplitude $A_\sigma$ in threshold region.
Consider this point in more details.

Pseudoscalar amplitude $A_\pi$ naturally associates with the pion,
which in the chiral limit is a massless Goldstone particle. In
both  regularizations under consideration we can  define a pion
propagator as a pole term of $A_\pi$, which corresponds to the
leading singularity of  pseudoscalar amplitude:
\begin{equation}
A_\pi^{pole}(p)=\frac{1}{4n_cI_0(0)p^2}, \label{Appole}
\end{equation}
where $I_0(0)$ is defined by equation (\ref{I0ARzero}) for
dimensionally-analytical regularization and by (\ref{I0cutzero})
for 4-dimensional cutoff.

 For the scalar amplitude the situation is different.
In both regularizations function $I_0(p^2)$  possesses a cut which
 originates in the point $p^2=4m^2$. Nevertheless, for 4-dimensional cutoff
 it is possible to define a scalar sigma-meson propagator as
\begin{equation}
A_\sigma^{pole}(p)=\frac{1}{4n_cI_0(4m^2)(4m^2-p^2)},
\label{Aspole}
\end{equation}
since
$$
I_0^{FDC}(4m^2)=\frac{i}{(4\pi)^2}\Bigl[\log\frac{\Lambda^2+m^2}{m^2}+
\frac{\Lambda}{m}\arctan\frac{m}{\Lambda}\Bigr]
$$
is a finite quantity. But for dimensionally-analytical
regularization this quantity  is finite only at $\xi<-1/2$:
$$
I_0^{DAR}(4m^2)\vert_{\xi<-1/2}=-\frac{i}{8gn_cm^2}
\frac{\xi}{1+2\xi}.
$$
For an interpretation of the sigma-meson as a particle in the NJL
model with dimensionally-analytical regularization we can do the
following trick: since in the region $-1<\xi<-1/2$ integral $I_0$
converges we use the above value in the point $p^2=4m^2$ as a
foundation for an analytical continuation of the pole part of the
amplitude on parameter $\xi$ to the physical region $0<\xi<1$.
Then the sigma-meson propagator for dimensionally-analytical
regularization would be
\begin{equation}
(A_\sigma^{pole}(p))^{DAR}= \frac{2igm^2(1+2\xi)}{(4m^2-p^2)\xi}.
\label{AspoleAR}
\end{equation}
This expression was used for a calculation of the sigma-meson
contribution in chiral condensate in work  \cite{JaRo}. Surely,
such procedure of definition of sigma-meson propagator seems to be
a somewhat artificial. A more consistent procedure is a separation
of a leading singular part of amplitude in the region of physical
values of regularization parameter $\xi$.

For the pseudoscalar amplitude the separation of leading
singularity near the point $p^2=0$ leads to same result
(\ref{Appole}), i.e. the pion in dimensionally-analytical
regularization possesses all properties of usual observable
particle. For the scalar amplitude it is not so.
 At $p^2\rightarrow 4m^2$ in region $0<\xi<1$:
$$
I_0^{DAR}\cong
\frac{i\sqrt{\pi}\Gamma(\xi+1/2)}{16gn_cm^2\Gamma(\xi)}\cdot
\Biggl(\frac{4m^2}{4m^2-p^2}\Biggr)^{\xi+1/2},
$$
and, correspondingly, the leading singularity ({\it LS}), i.e. a
leading term in an expansion on powers of $4m^2-p^2$ is the
expression
\begin{equation}
(A_\sigma^{LS})^{DAR}\cong
-\frac{ig\Gamma(\xi)}{\sqrt{\pi}\Gamma(\xi+1/2)}\cdot
\Biggl(\frac{4m^2}{4m^2-p^2}\Biggr)^{1/2-\xi}. \label{AsLS}
\end{equation}
Thus, the leading singularity of  scalar amplitude in the model
with dimensionally-analytical regularization is of  the
principally different type  in comparison with the cutoff model.
Instead  of the pole term, which can be naturally interpret as
sigma-particle propagator, we obtain in dimensionally-analytical
regularization the power behavior  which depends on the
regularization parameter $\xi$. Moreover,  due to formula
$$
F(3/2, 1; 3/2; \frac{p^2}{4m^2})=\frac{4m^2}{4m^2-p^2}
$$
 at  $\xi=1/2$ a cancellation of contributions
  into two-particle amplitude takes
place (see also \cite{Ro3}). At this parameter value ($\xi=1/2$)
we obtain for the amplitudes extremely simple expressions:
\begin{equation}
A_\pi=ig-\frac{4igm^2}{p^2},\;\;\;A_\sigma=-ig \label{xi1/2},
\end{equation}
i.e. at $\xi=1/2$ the scalar amplitude has no singularity --
sigma-meson disappears! We emphasize, that result (\ref{xi1/2}) is
an exact consequence of formulae (\ref{A_sigma}), (\ref{A_pi}) and
(\ref{I0AR}) without any approximation type of above
leading-singularity approximation.

Thus, we come to the conclusion, that for dimensionally-analytical
regularization at physical values of parameters the scalar
amplitude $A_\sigma$ does not possesses a pole term, which can be
interpret as a physical scalar meson.

\section{Meson contributions in chiral condensate and in quark propagator}

First-order equations of the mean-field expansion (see
\cite{JaRo}) define corrections to quark propagator. First-order
mass operator $\Sigma^{(1)}=S^{-1}_0\star S^{(1)}\star S^{-1}_0$,
where $S^{(1)}$ is a first-order correction to quark propagator,
is defined in $x$-space by equation (see \ref{JaRo})
\begin{equation}
\Sigma^{(1)}(x) =
S^{(0)}(x)A_\sigma(x)+3S^{(0)}(-x)A_\pi(x)+ig\delta(x)\tr
S^{(1)}(0). \label{Sigma}
\end{equation}
Introducing dimensionless first-order mass functions $a^{(1)}$ and
$b^{(1)}$:
\begin{equation}
\Sigma^{(1)}\equiv a^{(1)}\hat p-b^{(1)}m, \label{a1b1}
\end{equation}
and also defining the first-order condensate
\begin{equation}
\chi^{(1)}=i\mbox{tr}S^{(1)}(0) \label{chi1}
\end{equation}
and a ratio of the first-order condensate to the leading-order
condensate
$$
r\equiv \frac{\chi^{(1)}}{\chi^{(0)}},
$$
we obtain from  (\ref{Sigma}) the expressions for $a^{(1)}$ and
$b^{(1)}$ in momentum space:
\begin{equation}
p^2a^{(1)}(p^2)=\int d\tilde q\frac{ p^2-(pq)
}{m^2-(p-q)^2}[A_\sigma(q)-3A_\pi(q)], \label{a1}
\end{equation}
\begin{equation}
b^{(1)}(p^2)=r-\int \frac{d\tilde
q}{m^2-(p-q)^2}[A_\sigma(q)+3A_\pi(q)]. \label{b1}
\end{equation}
It follows from equations  (\ref{a1}) and (\ref{b1}), that the
corrections to quark propagator consist of two parts: pion
correction (due to pseudoscalar amplitude $A_\pi$) and
contribution due to scalar amplitude $A_\sigma$:
$a^{(1)}=a^{(1)}_\pi+a^{(1)}_\sigma;\;\;b^{(1)}=b^{(1)}_\pi+b^{(1)}_\sigma.$

For the ratio of the first-order condensate (\ref{chi1}) to the
leading-order condensate (\ref{chi0}) we obtain the formula
\begin{equation}
r=-\frac{g\chi^{(1)}}{m}=- 8ign_c\int d\tilde p
\frac{2p^2a_1-(m^2+p^2)b_1}{(m^2-p^2)^2}. \label{r}
\end{equation}
Corrections to quark mass can be found after the calculation the
condensate corrections. Inverse quark propagator is
$$
S^{-1}=m-\hat{p}-\Sigma^{(1)}=b(p^2)-a(p^2)\hat{p}=(1+b^{(1)})m-(1+a^{(1)})\hat{p}.
$$
Suppose the  propagator  has a pole in point  $p^2=m^2_r$, which
corresponds to a particle with mass $m_r$. Then
$$
b(m^2_r)=m_ra(m^2_r).
$$
Since  $a^{(1)}$ and $b^{(1)}$ are small additions, we can to
expand
 $a^{(1)}(m^2_r)$ and
$b^{(1)}(m^2_r)$ near the point  $m$ and to obtain the formula for
the quark-mass correction $ \delta m\equiv m_r-m$:
\begin{equation}
\frac{\delta m}{m}\cong b^{(1)}(m^2)-a^{(1)}(m^2). \label{delta m}
\end{equation}

\subsection{Pion contribution}

The pion contribution to quark propagator is defined by formulae
\begin{equation}
p^2a_{\pi}^{(1)}(p^2)=-3\int d\tilde q\frac{ p^2-(pq)
}{m^2-(p-q)^2}A_\pi(q), \label{a1pi}
\end{equation}
\begin{equation}
b_{\pi}^{(1)}(p^2)=r_\pi-3\int \frac{d\tilde q
}{m^2-(p-q)^2}A_\pi(q). \label{b1pi}
\end{equation}

For the calculation we shall use the pole approximation
(\ref{Appole}). The calculation reduces to calculations of
integrals
\begin{equation}
I_0(p^2; m^2, \mu^2)=\int \frac{d\tilde
q}{(m^2-(p-q)^2)(\mu^2-q^2)}, \label{I0mu}
\end{equation}
\begin{equation}
I_\nu(p^2; m^2, \mu^2)=\int \frac{q_\nu d\tilde
q}{(m^2-(p-q)^2)(\mu^2-q^2)} \label{Inu}
\end{equation}
 at $\mu^2\rightarrow 0$.

These integrals are calculated with above rules (see Sections 2
and 3). The pion contribution to first-order condensate is
calculated by formula (\ref{r}). Integral  can be calculated in a
closed form, and the result is the  very simple expression:
\begin{equation}
(r_\pi)^{DAR}=\frac{3}{8n_c\xi} \label{rpiAR}.
\end{equation}
(See also \cite{JaRo}, where this result have been obtained by a
slightly different method.)

To calculate the pion contribution for 4-dimensional cutoff we use
equations  (\ref{Appole}) and (\ref{I0cutzero}). Further the pion
contribution in condensate is calculated with formula (\ref{r}).
For 4-dimensional cutoff the result for $r_\pi$ is not described
by a simple formula, as for dimensionally-analytical
regularization. Nevertheless, the computation has no any troubles.
Note, that whereas in dimensionally-analytical regularization
$r_\pi$ is a function of regularization parameter $\xi$, in
4-dimensional cutoff this quantity is a function of ratio
$x\equiv\Lambda^2/m^2$:
$$
(r_\pi)^{FDC} = r_\pi(\Lambda^2/m^2).
$$
As  examples we give values of $(r_\pi)^{FDC}$ for two
characteristic values of this ratio\footnote{All  values given
here  correspond to physical value of colours $n_c=3$.}. At $x=3$
(this value corresponds to value $c^{(0)}=-210$MeV of the
leading-order condensate) the computation gives
$(r_\pi)^{FDC}=-0.272$. At $x=19$ (this value corresponds to value
$c^{(0)}=-250$MeV of the leading-order condensate) the computation
gives $(r_\pi)^{FDC}=-0.183$.

Let turn to the pion contribution   in quark mass. To calculate
this contribution we apply equation (\ref{delta m}).
 As a result we obtain (with taking into account gap
equation (\ref{gapARM}) and equation (\ref{rpiAR})):
\begin{equation}
\biggl(\frac{\delta m_{(\pi)} }{m}\biggr)^{DAR}
=(r_\pi)^{DAR}-\frac{3}{8n_c\xi}=0, \label{deltapiAR}
\end{equation}
i.e. for the dimensionally-analytical regularization the pion
correction to quark mass equal zero.

For 4-dimensional cutoff the pion correction to quark mass is
\begin{equation}
\biggl(\frac{\delta m_{(\pi)}
}{m}\biggr)^{FDC}=(r_\pi)^{FDC}+\frac{3}{n_c}h_\pi(\Lambda^2/m^2),
\label{deltapicut}
\end{equation}
where
$$
h_\pi(x)=\frac{\log(1+x)}{8[\log(1+x)-\frac{x}{1+x}]}.
$$
Signs of $(r_\pi)^{FDC}(x)$ and  $h_\pi(x)$ are opposite, and
their contributions in $\delta m$ are mutually cancelled.
Moreover, as for the dimensionally-analytical regularization, the
pion correction to quark mass equal zero. We obtain this result,
unlike to the exact result (\ref{deltapiAR}) of
dimensionally-analytical regularization, by computations in a
framework of given accuracy of inputs. Such coincidence of results
in both regularizations suggests an idea that the  zero value of
the pion correction to quark mass is the
regularization-independent fact of NJL model.

\subsection{Scalar contribution}
Consider a contribution of scalar amplitude $A_\sigma$ in
condensate and quark mass. In correspondence with  (\ref{b1}) and
(\ref{a1}) we have
\begin{equation}
 p^2a_{\sigma}^{(1)}=\int d\tilde q
\frac{ p^2-(pq)}{m^2-(p-q)^2}A_\sigma(q), \label{a1s}
\end{equation}
\begin{equation}
b_{\sigma}^{(1)}=r_\sigma-\int \frac{d\tilde q
}{m^2-(p-q)^2}A_\sigma(q). \label{b1s}
\end{equation}

To calculate this contribution we  use the leading-singularity
approximation:
$$
A_\sigma^{LS}=\frac{1}{4n_c(4m^2-p^2)I_0(p^2)\vert_{p^2\rightarrow
4m^2} }.
$$
For the dimensionally-analytical regularization this approximation
is described by equation (\ref{AsLS}).
From equation (\ref{r})
we obtain the quantity $r_\sigma$.
A computation gives us the following values for
sigma-contribution: at $\xi=0.25$ we obtain
$(r_\sigma)^{DAR}=-0.033$; at $\xi=0.4$ we obtain
$(r_\sigma)^{DAR}=-0.01$. As one can see, the sigma-contribution
is small in comparison of the pion contribution and possesses the
opposite sign, i.e. it decrease the common contribution\footnote{
 Note, that this result is qualitatively the same as
 result of work \cite{JaRo}, in which has been used a pole approximation
 for $A_\sigma$. Thus, all conclusions of work   \cite{JaRo} about
 the part of the meson contributions stand also for the more exact
  leading-singularity approximation, which is used in present work}.

For the 4-dimensional cutoff the leading-singularity approximation for
 $A_\sigma$ coincides with the pole approximation(\ref{Aspole}).
Then $r_\sigma$ is calculated by equation (\ref{r}). This
quantity, as $r_\pi$, for the 4-dimensional cutoff is a function
of
  $x\equiv\Lambda^2/m^2$:
$$
(r_\sigma)^{FDC}=r_\sigma(\Lambda^2/m^2).
$$
At $x=3$ we obtain $(r_\sigma)^{FDC} =-0.007$. At $x=19$ we obtain
$(r_\sigma)^{FDC} =-0.116$. In contrast to the
dimensionally-analytical regularization, the sign of
sigma-contribution for the 4-dimensional cutoff is the same
 as for pion contribution.

A sigma-correction to quark mass for dimensionally-analytical
regularization is given by formula
\begin{equation}
\biggl(\frac{\delta m_{(\sigma)}}{m}\biggr)^{DAR}=
 (r_\sigma)^{DAR}-\frac{\cos
\pi\xi}{4^{1+\xi} n_c\pi (1/2-\xi)} \label{deltasigmaAR}
\end{equation}
and attains: at $\xi=0.25:\;\;\delta m_{(\sigma)}^{DAR} =-0.086m$,
and at $\xi=0.4:\;\;\delta m_{(\sigma)}^{DAR} =-0.056m$. Since a
pion correction to quark  mass in this regularization equals zero
(see above), these values are  full corrections  to quark mass for
dimensionally-analytical regularization.

For the 4-dimensional cutoff the sigma-correction to quark-mass is
\begin{equation}
\biggl(\frac{\delta m_{(\sigma)} }{m}\biggr)^{FDC}
=(r_\sigma)^{FDC}-\frac{1}{n_c}h_\sigma(\Lambda^2/m^2),
\label{deltasigmacut}
\end{equation}
where
$$
h_\sigma(x)=\frac{4\log(1+x/4)-\log(1+x)}{8[\log(1+x)+\sqrt{x}\arctan\sqrt{
\frac{1}{x} }]}.
$$
At $x=3$:  $\;\;\delta m_{(\sigma)}^{FDC} =-0.022m$; at
$x=19$: $\;\;\delta m_{(\sigma)}^{FDC}= -0.158m$.   \\

In conclusion of this Section let consider an issue on an accuracy
of above calculations. A principal  approximation of above
calculations is the leading-singularity approximation.
 Let consider a part of other terms. To estimate their part
 for dimensionally-analytical
 regularization let use the simple expressions of amplitudes at
 $\xi=1/2$ (see (\ref{xi1/2})).  Remind these expressions are exact.
 A calculation with formulae (\ref{a1})--(\ref{r}) demonstrates that
 the contributions of non-pole terms in chiral condensate equal zero.
 Since the values of parameter $\xi$ are near this point (see below,
 Table 3), we can maintain, that at $\xi\neq 1/2$ their contributions
 are also small in comparison with the main pole contribution.

For 4-dimensional cutoff the calculations with exact formulae
 (\ref{A_sigma})-(\ref{A_pi}) for the amplitudes also demonstrate,
 that the leading-singularity approximation
 (pole approximation in the case) gives the main contribution in
 condensate. So, at $x=3$ the calculation with the exact formulae
 (\ref{A_sigma})-(\ref{A_pi}) gives for the pion contribution
 $r_\pi=-0.267$, i.e., differs from the result of pole approximation
 (see  Subsection 3.1) less then on 2\%. For sigma-contribution
 the difference is more significant: the calculation with the exact
 formulae gives $r_\sigma=-0.031$, but since this contribution is much
 less in comparison with the pion contribution, this difference again
 practically does not affect to final result.

\section{Improved model parameters}

The condensate and the quark-mass corrections, which were
calculated in preceding Section, allow us to specify parameters of
the SU(2)-NJL model. We modify a formula for the condensate as
follows:
\begin{equation}
\chi=\chi^{(0)}+\chi^{(1)}=- \frac{m}{g}(1+r).
\label{condimproved}
\end{equation}
The formula for $f_\pi$ (see (\ref{fNJL})) stays the same, since
corrections to amplitudes generate in the next (second) order of
mean-field expansion. The quark mass is the mass $m_r$:
$$
m_r=m+\delta m,
$$
where $\delta m$ is defined by equation (\ref{delta m}). Values of
the model parameters at $n_c=3$ for this improved choice  are
given in Tables 3 and 4.

\begin{center}
\begin{tabular}{|l|l|l|l|l|l|}\hline
$c$ (MeV)
& $m_r$ (MeV)& $\xi$ & $\kappa=3gm^2/2\pi^2$ \\
\hline
\  -210    & \  339 & \ 0.432  & \  0.486\\
\  -220   &  \  336 & \ 0.385 & \ 0.434 \\
\  -230   &  \  333 & \  0.346 & \ 0.387 \\
\  -240  & \ 330  &\  0.312 & \  0.334 \\
\  -250   &  \ 328   & \  0.284 & \ 0.316 \\

\hline

\end{tabular}
\vspace{1cm}

{\small Table 3. Model parameters with first-order corrections
(dimensionally-analytical regularization): chiral condensate $c$,
quark mass $m_r$, regularization parameter $\xi$ and
dimensionless coupling $\kappa$.}
\end{center}
\vspace{1.5cm}

\begin{center}
\begin{tabular}{|l|l|l|l|l|l|}\hline
 $c$ (MeV)
 & $m_r$ (MeV)& $\Lambda$ (MeV)& $\kappa_\Lambda=3g\Lambda^2/2\pi^2$  \\
 \hline
\  -240 &   \ 310  & \  785 & \  1.501 \\
\  -250 &   \ 283  & \ 819 & \ 1.408 \\

 \hline

\end{tabular}
\vspace{1cm}

{\small Table 4. Model parameters with first-order corrections
(4-dimensional cutoff): chiral condensate $c$, quark mass
$m_r$, regularization parameter $\Lambda$  and dimensionless
coupling $\kappa_\Lambda$.}
\end{center}

Table 4 does not contain the parameter values at $c=-210$ MeV,
$c=-220$ MeV and  $c=-230$  MeV. These values are absent due to
following reason: the system of equations (\ref{condimproved}),
(\ref{fcut}) and (\ref{gapcut}), which determines these
parameters, has no solution at
 $f_\pi=93$ MeV and at $\vert c \vert \leq 230$ MeV. There is very important
 circumstance -- for 4-dimensional cutoff the meson contributions
 can destabilize the NJL model. Though these contributions are
 relatively small (they do not exceed 25\% from the leading
 contribution), but their opposite sign leads to a non-stability
 of all the system. The situation is very similar to that of
 pointed in work \cite{Kleinert}. Note, that  increasing a number
of flavours, i.e. for the U($n_f$)-NJL model ($n_f$ is a number of
flavours), the situation takes a turn for the worse,  because a
main pseudoscalar contribution is proportional to $n_f$. At that
for dimensionally-analytical regularization the situation is
principally different: due to the coincidence of sign of the meson
contributions with the sign of leading contribution  in condensate
for this regularization a stabilization of the model takes place.
It is clearly seen from Table 3 --  values of
 regularization parameter $\xi$ increase in comparison with
 corresponding leading-order values (see Table 1), i.e. shift to
 a region of stability of model, where these meson contributions
 decrease.

\section*{Conclusion}

The results of present work demonstrate that the NJL model with
dimensionally-analytical regularization  essentially differs from
the NJL model with 4-dimensional cutoff at least in two aspects.

Firstly, there is the  different behavior of scalar amplitude in
threshold region. For the 4-dimensional cutoffit is possible to
separate near the threshold  a pole term, which is usually
associated with a scalar particle -- sigma-meson (note, however,
that reasoning doubts in such interpretation have been stated as
early as in founder's work \cite{NJL}). For the
dimensionally-analytical regularization the singularity of scalar
amplitude is not of pole type at physical values of regularization
parameter. This fact, even if does not exclude entirely, makes its
interpretation as a physical particle to be awkward.

But much more principal thing  is the different behavior of these
models with respect to quantum fluctuations caused by meson
contributions in chiral condensate. As it follows from results of
Sections 3 and 4, the NJL model with dimensionally-analytical
regularization is stable with respect to these fluctuations,
whereas for the NJL model with 4-dimensional cutoff the meson
contributions can lead to destabilization. Surely, a number of
physical applications of the NJL model are connected exclusively
with the leading order of mean-field expansion (mean-field
approximation), for which the possibility of such destabilization
can be simply ignored. On the other hand, some physical
applications of the NJL model exist, that connected with
multi-quark functions (such as pion-pion scattering, baryons
etc.). For these applications the  neglecting by the meson
contributions in quark propagator is certainly non-correct from
the point of view of the mean-field expansion, and, consequently,
the stability of basic model parameters with respect to these
contributions becomes a determinative significance.

We  thank Prof. S.A. Hajiev and Dr. M.L. Nekrasov for fruitful
discussions. One of\\ authors (R.G.J) would like to take the
opportunity to thank BSU's chancellor Prof. A.M. Maharramov for
attention and support.


\begin{thebibliography}{99}
\bibitem{NJL} Y. Nambu and G. Jona-Lasinio: Phys.Rev. {\bf 122}
(1961) 345
\bibitem{QNJL}
T. Eguchi and H. Sugawara: Phys.Rev.  D {\bf 10} (1974) 4257; \\
K. Kikkawa: Prog.Theor.Phys. {\bf 56} (1976) 947;\\
 H. Kleinert: in  "Understanding the Fundamental Constituents of
Matter", ed. A. Zichichi, Plenum Press, N.Y., 1978, p.289;\\
D. Ebert and M.K. Volkov: Z.Phys. C {\bf 16} (1983) 305
\bibitem{kle} S.P. Klevansky: Rev.Mod.Phys. {\bf 64} (1992)
649
\bibitem{HatKun}
T. Hatsuda and T. Kunihiro: Phys.Reports {\bf 247} (1994) 221
\bibitem{Nikolov}
E.N. Nikolov et al.: Nucl.Phys. {\bf A608} (1996) 411
\bibitem{Quack}
E. Quack and S.P. Klevansky: Phys.Rev. C {\bf 49} (1994) 3283\\
D. Ebert, M. Nagy and M.K. Volkov: Yad.Fiz. {\bf 59} (1996) 149
\bibitem{Oertel}
 M. Oertel, M. Buballa and J.
Wambach: Nucl.Phys. {\bf A676} (2000) 247
\bibitem{Krewald}
 S. Krewald and K. Nakayama: Annals
of Phys. {\bf 216} (1992) 201
\bibitem{JaRo}
R.G. Jafarov and V.E. Rochev : Centr.Eur.J.of Phys. {\bf 2} (2004)
367  (hep-ph/0311339)
\bibitem{Ochs}
 W. Ochs:  Plenary talk at Hadron 03 (Aschaffenburg, Germany),
  hep-ph/0311144
\bibitem{Kleinert}
H. Kleinert and B. Van den Bossche: Phys.Lett. {\bf B474} (2000)
336
\bibitem{Ripka}
 E. Babaev: Phys.Rev. {\bf D62} (2000) 074020;\\
 G. Ripka:  Nucl.Phys. {\bf A683} (2001) 463
\bibitem{Ro1}
V.E. Rochev : J.Phys. A: Math.Gen. {\bf 30} (1997) 3671\\
V.E.  Rochev and P.A. Saponov : Int.J.Mod.Phys. {\bf A13}
(1998) 3649 \\
V.E. Rochev : J.Phys. A: Math.Gen. {\bf 33} (2000) 7379
\bibitem{Ro3}
V.E. Rochev : Talk given at XVII International Workshop on High
Energy Physics and Quantum Field Theory (QFTHEP'2003),
 Samara--Saratov, 4-11 Sept. 2003 (hep-ph/0312004)



\end{thebibliography}
\end{document}